\documentclass[a4paper,10pt]{article}
\usepackage{graphicx}
\usepackage{amsmath}
\usepackage{citesort}
\usepackage{amsfonts}
\usepackage{amssymb}
\usepackage{amsmath}
\usepackage{latexsym}
\usepackage{color}
\usepackage{ntheorem}
\usepackage{braket}
   \newcommand{\lambdahere}{\lambda}%
\newcommand{\oyinf}{O(|y|^{\infty})}%
\newcommand{\orinf}{O(r^{\infty})}%
\newcommand{\olt}[1]{\overline{\tilde #1}{}}%
\newcommand{\old}[1]{\overline{\delta #1}{}}%
\newcommand{\olc}[1]{\overline{\check #1}{}}%
\newcounter{mnotecount}[section]

\renewcommand{\themnotecount}{\thesection.\arabic{mnotecount}}

\newtheorem{theorem}{Theorem}[section]

\newtheorem{lemma}[theorem]{\sc Lemma\rm}

\newtheorem{Proposition}[theorem]{\sc Proposition\rm}

\newtheorem{Remark}[theorem]{\sc Remark\rm}

\newcommand{\ol}[1]{\overline{#1}{}}

\newcommand{\jlcax}[1]{}
\newcommand{\eean}{\nonumber\end{eqnarray}}

\newcommand{\kk}[1]{}%{\mnote{{\bf If we consider the KK case:} #1}}

\newcommand{\beq}{\begin{equation}}

\newcommand{\FS}       %{F_1} %
                  {F}
\newcommand{\HS} %{F_2}
       {H_{\mbox{\scriptsize volume}}}

{\ptc{this should be removed in the oberwolfach version}}%

\newcommand{\eeal}[1]{\label{#1}\end{eqnarray}}
\newcommand{\bed}{\begin{deqarr}}
\newcommand{\eed}{\end{deqarr}}
\newcommand{\bedl}[1]{\begin{deqarr}\label{#1}}
\newcommand{\eedl}[2]{\arrlabel{#1}\label{#2}\end{deqarr}}

\newcommand{\mcO}{{\mycal O}}
\newcommand{\mcU}{{\mycal U}}

\newcommand{\bel}[1]{\begin{equation}\label{#1}}
\newcommand{\bea}{\begin{eqnarray}}
\newcommand{\bean}{\begin{eqnarray}\nonumber}
\newcommand{\beal}[1]{\begin{eqnarray}\label{#1}}
\newcommand{\eea}{\end{eqnarray}}

\newcommand{\qed}{\hfill $\Box$ \medskip}
\newcommand{\proof}{\noindent {\sc Proof:\ }}
\newcommand{\be}{\begin{equation}}
\newcommand{\eeq}{\end{equation}}
\newcommand{\ee}{\end{equation}}
\newcommand{\beqa}{\begin{eqnarray}}
\newcommand{\eeqa}{\end{eqnarray}}
\newcommand{\beqan}{\begin{eqnarray*}}
\newcommand{\eeqan}{\end{eqnarray*}}
\newcommand{\ba}{\begin{array}}
\newcommand{\ea}{\end{array}}

\newcommand{\mcM}{{\mycal M}}
\newcommand{\mcD}{{\mycal D}}

\newcommand{\scri}{{\mycal I}}%
\newcommand{\scrim}{\scri^{-}}%

\newcommand{\mnote}[1]%{}
{\protect{\stepcounter{mnotecount}}$^{\mbox{\footnotesize
$%\!\!\!\!\!\!\,
\bullet$\themnotecount}}$ \marginpar{%\color{red}%
\raggedright\tiny\em
$\!\!\!\!\!\!\,\bullet$\themnotecount: #1} }

\newcommand{\warn}[1]%{}%{}
{\protect{\stepcounter{mnotecount}}$^{\mbox{\footnotesize
$%\!\!\!\!\!\!\,
\bullet$\themnotecount}}$ \marginpar{%\color{red}%
\raggedright\tiny\em
$\!\!\!\!\!\!\,\bullet$\themnotecount: {\bf Warning:} #1} }

\newcommand{\R}{\mathbb R}

\newcommand{\N}{\mathbb N}

\newcommand{\eq}[1]{(\ref{#1})}

\newcommand{\ptc}[1]{\mnote{{\bf ptc:}#1}}

\newcommand{\beqar}{\begin{deqarr}}
\newcommand{\eeqar}{\end{deqarr}}

\newcommand{\beaa}{\begin{eqnarray*}}
\newcommand{\eeaa}{\end{eqnarray*}}

\DeclareFontFamily{OT1}{rsfs}{}
\DeclareFontShape{OT1}{rsfs}{m}{n}{ <-7> rsfs5 <7-10> rsfs7 <10-> rsfs10}{}
\DeclareMathAlphabet{\mycal}{OT1}{rsfs}{m}{n}

{\catcode `\@=11 \global\let\AddToReset=\@addtoreset}
\AddToReset{equation}{section}

{\catcode `\@=11 \global\let\AddToReset=\@addtoreset}
\AddToReset{figure}{section}

{\catcode `\@=11 \global\let\AddToReset=\@addtoreset}
\AddToReset{table}{section}

\newcommand{\nottmathrmD}{\mathrm{D}}
\newcommand{\ttilde}[1]{\tilde{\tilde{#1}}{}}

\begin{document}

%\pagestyle{myheadings}
%\markright{The many ways of the characteristic Cauchy problem}

\title{Solutions of the vacuum Einstein equations with initial data on past null infinity%
\thanks{Preprint UWThPh-2013-13.  Supported in part by the  Austrian Science Fund (FWF): P 24170-N16.}}
\author{
Piotr T. Chru\'sciel%{}
\thanks{Email  {Piotr.Chrusciel@univie.ac.at}, URL {
http://homepage.univie.ac.at/piotr.chrusciel}}
%\\ Helmut Friedrich\thanks{AEI Golm; Email hef@aei.mpg.de}
\\
Tim-Torben Paetz%
\thanks{Email  Tim-Torben.Paetz@univie.ac.at}  \\  Gravitational Physics, University of Vienna\\
 Boltzmanngasse 5, 1090 Vienna, Austria   }
\maketitle

\vspace{-0.2em}

\begin{abstract}
We prove existence of vacuum space-times with  freely prescribable cone-smooth
initial data on past null infinity.
\end{abstract}

\noindent
\hspace{2.1em} PACS: 04.20.Ex, 04.20.Ha

\tableofcontents

\section{Introduction}

A question of interest in general relativity is the construction of large classes of space-times with controlled global properties. A flagship example of this line of enquiries is the Christodoulou-Klainerman theorem~\cite{Ch-KL} of nonlinear stability of Minkowski space-time. Because this theorem carries only limited information on the asymptotic behaviour of the resulting gravitational fields, and applies only to near-Minkowskian configurations
in any case, it is of interest to construct space-times with better understood global properties. One way of doing this is to carry out the construction starting from initial data at the future null cone, $\scrim$, of past timelike infinity $i^-$. An approach to this has been presented in~\cite{FriedrichCMP86}, but an existence theorem for the problem is still lacking.
The  purpose of this work is to fill this gap.

In order to present our result some terminology and notation is needed: Let $C_O$ denote the (future) light-cone of the origin $O$ in Minkowski space-time  (throughout this work, by ``light-cone of a point $O$" we mean the subset of a spacetime $\mcM$ covered by future directed null geodesics issued from $O$).   Let, in manifestly flat coordinates $y^\mu$, $\ell=\partial_0+(y^i/|\vec y|) \partial_i$ denote the field of null tangents to $C_O$.
Let $\tilde d_{\alpha\beta\gamma\delta}$ be a tensor with algebraic symmetries of the Weyl tensor and with vanishing $\eta$-traces, where $\eta$ denotes the Minkowski metric.  Let $\varsigma$ be the pull-back of
$$
 \tilde d_{\alpha\beta\gamma\delta}\ell^\alpha \ell^\gamma
$$
to $C_O\setminus \{O\}$.   Let, finally, $\varsigma_{ab}$ denote the components of $\varsigma$ in a frame
parallel-propagated along the generators of $C_O$. We  prove the following:

\begin{theorem}
 \label{T19XII12.1}
Let $C_O$ be the light-cone of the origin $O$ in Minkowski space-time.
For any   $\varsigma$ as above there exists a neighborhood $\mcO$ of $O$, a smooth metric $g$ and a
{smooth}
function $\Theta$
such that $C_O$ is the light-cone of $O$ for $g$, $\Theta$ vanishes on $C_O$, with $\nabla \Theta$ nonzero on $\dot J^+(O)\cap \mcO\setminus \{O\}$,
the function $\Theta$ has no zeros on $\mcO\cap I^+(O)$, and the metric  $\Theta^{-2}g$ satisfies the vacuum Einstein equations there. Further, the tensor field
$$
  d_{\alpha\beta\gamma\delta} := \Theta^{-1} C_{\alpha\beta\gamma\delta}
   \;,
$$
where  $C_{\alpha\beta\gamma\delta}$ is the Weyl tensor of $g$,
extends smoothly across $\{\Theta=0\}$, and  $\varsigma_{ab} $ are the frame components, in a $g$-parallel-propagated frame, of  the pull-back to $C_O$ of $d_{\alpha\beta\gamma\delta}\ell^\alpha \ell^\gamma$.     The solution is unique up to isometry.
\end{theorem}

\subsection{Strategy of the proof}

The starting point of our analysis are the conformal field equations of Friedrich. The task consists of constructing initial data, for those equations, which arise as the restriction to the future light-cone $\scrim$ of past  {timelike}
infinity $i^-$ of tensors which are smooth in the unphysical space-time. We then use a  system of conformally invariant wave equations of~\cite{TimConformal} to obtain a space-time with a metric solving the vacuum Einstein equations to the future of $i^-$.

Now, some of Friedrich's conformal equations involve only derivatives tangent to $\scrim$, and have therefore the character of \emph{constraint equations}. Those equations form a set of PDEs with a specific hierarchical structure, so that solutions can be obtained by integrating ODEs along the generators of $\scrim$. This implies that the constraint equations can be solved in a straightforward way in coordinates adapted to $\scrim$ in terms of a subset of the fields on the light-cone.
However, there arise  serious  difficulties when attempting to show that  solutions of the conformal constraint equations can be realized by  smooth space-time tensors.
These difficulties lie  at the heart of the problem at hand.  To be able to handle this issue, we note that $\varsigma$ determines the \emph{null data} of~\cite{FriedrichNullData}. These null data are used in~\cite{FriedrichNullData} to construct smooth tensor fields  satisfying Friedrich's  equations
up to  terms which decay faster than any power of the Euclidean coordinate distance  from $i^-$, similarly for their derivatives of any order; such error terms are said to be  $\oyinf$. For fields on the light-cone, the notation $\orinf$ is defined similarly, where $r$ is an affine distance from the vertex along the generators, with derivatives only in directions tangent to the light-cone.
In particular the \emph{approximate solution} so obtained solves the constraint equations up to error terms of order $\orinf$.   Using a comparison argument, we show that the approximate fields differ, on $C_O$, from  the exact solution of the constraints by terms which are  $\orinf$. But tensor fields on the light-cone which decay to infinite order in adapted coordinates arise from smooth tensors in space-time, which implies that the solution of the constraint equations arises indeed from a smooth tensor in space-time. As already indicated, this is what is needed to be able to apply the existence theorems for systems of wave equations in~\cite{Dossa97}, provided such a system is at disposal.
This last element of our proof is provided by the system of wave equations of~\cite{TimConformal}, and the results on
propagation of   constraints for this system established there.%
\footnote{Compare~\cite{DossaAHP}, where a system based on the equations of Choquet-Bruhat and Novello~\cite{ChBNovello} is used.}

\section{From approximate solutions to solutions}
 \label{s13X12.1}

Recall Friedrich's  system of conformally-regular equations (see~\cite{Friedrich:tuebingen} and references therein)
%\tim{good reference?}
%
\begin{eqnarray}
 && \nabla_{\rho} d_{\mu\nu\sigma}{}^{\rho} =0
  \;,
 \label{mconf1}
\\
 && \nabla_{\mu}L_{\nu\sigma} - \nabla_{\nu}L_{\mu\sigma} = \nabla_{\rho}\Theta \, d_{\nu\mu\sigma}{}^{\rho}
  \;,
 \label{mconf2}
\\
 && \nabla_{\mu}\nabla_{\nu}\Theta = -\Theta L_{\mu\nu} +  s  g_{\mu\nu}
  \;,
 \label{mconf3}
\\
 && \nabla_{\mu}  s = - L_{\mu\nu}\nabla^{\nu}\Theta
  \;,
 \label{mconf4}
\\
 && 2\Theta  s -  \nabla_{\mu}\Theta\nabla^{\mu}\Theta = 0
  \;,
 \label{mconf5}
\\
 &&  R_{\mu\nu\sigma}{}^{\kappa}[ g] = \Theta  d_{\mu\nu\sigma}{}^{\kappa} + 2\left( g_{\sigma[\mu} L_{\nu]}{}^{\kappa}  - \delta_{[\mu}{}^{\kappa} L_{\nu]\sigma} \right)
 \label{mconf6}
\;.
\end{eqnarray}
Here $\Theta$ is the conformal factor relating the physical metric $\tilde g_{\mu\nu}$ with the unphysical metric $g_{\mu\nu}=\Theta^{-2} \tilde g_{\mu\nu}$, the fields $d_{\mu\nu\sigma}{}^{\kappa}$ and $L_{\alpha\beta}$ encode the information about the unphysical Riemann tensor as made explicit in \eq{mconf6}, while the trace of \eq{mconf3} can be viewed as the definition of $s$.

We wish to construct  solutions of \eq{mconf1}-\eq{mconf6} with initial data on a light-cone $C_{i^-}$, emanating from a point $i^-$, with $\Theta$ vanishing on $C_{i^-}$ and with $s(i^{-})\ne 0$. (The actual value of $s(i^{-})$ can be changed by   constant rescalings of the conformal factor $\Theta$ and of the field $d_{\alpha\beta\gamma}{}^\delta$. For definiteness we will choose $s(i^{-})=-2$.)
As explained in~\cite{F1}, such solutions lead to   vacuum space-times, where past timelike infinity is the point $i^-$ and where past null infinity $\scri^-$  is $C_{i^-}\setminus  \{i^-\}$.

We will present two methods of doing this: while the second one is closely related to the classical one in~\cite{ChBNovello}, the advantage of the first one is that it allows in principle a larger class of initial data, see Remark~\ref{r6.12.2} below.

Let, then, a ``target metric" $\hat g$ be given and let the operator  $\hat\nabla$  denote its covariant derivative with associated
Christoffel symbols $\hat\Gamma^{\sigma}_{\alpha\beta}$. Set
\begin{equation}
 \label{10VI13.1}
 H^{\sigma} := g^{\alpha\beta}(\Gamma^ {\sigma}_{\alpha\beta} -\hat\Gamma^{\sigma}_{\alpha\beta} )
  \;.
\end{equation}
Consider  the system of wave equations which~\cite[Section~6.1]{TimConformal}
follows from \eq{mconf1}-\eq{mconf6}
when   $H^\sigma$ vanishes:
\begin{eqnarray}
  \Box^{(H)}_{ g} L_{\mu\nu}&=&  4 L_{\mu\kappa} L_{\nu}{}^{\kappa} -  g_{\mu\nu}| L|^2
  - 2C_{\mu\sigma\nu}{}^{\rho}  L_{\rho}{}^{\sigma}
 + \frac{1}{6}\nabla_{\mu}\nabla_{\nu} R
   \;,
 \label{weylwave1}
\\
 \Box _{ g}  s  &=& \Theta| L|^2 -\frac{1}{6}\nabla_{\kappa} R\,\nabla^{\kappa}\Theta  - \frac{1}{6} s  R
  \;,
 \label{weylwave2}
\\
  \Box _{ g}  \Theta &=& 4 s-\frac{1}{6} \Theta  R
  \;,
 \label{weylwave3}
\\
 \Box^{(H)}_g C_{\mu\nu\sigma\rho} &=&    C_{\mu\nu\alpha}{}^{ \kappa} C_{\sigma\rho\kappa}{}^{\alpha}  -4C_{\sigma\kappa[\mu}{}^{\alpha}  C_{\nu]\alpha \rho}{}^{ \kappa}     - 2C_{\sigma\rho\kappa[\mu} L_{\nu]}{}^{ \kappa}    - 2 C_{\mu\nu\kappa[\sigma} L_{\rho]}{}^{\kappa}
 \nonumber
\\
 &&
   -\nabla_{[\sigma}\xi_{\rho]\mu\nu}  -\nabla_{[\mu}\xi_{\nu]\sigma\rho} +  \frac{1}{3} R C_{\mu\nu\sigma\rho}
  \;,
 \label{weylwave4}
\\
 \Box^{(H)}_{ g}\xi_{\mu\nu\sigma} &=& 4 \xi_{\kappa\alpha[\nu} C_{\sigma]}{}^{\alpha}{}_{\mu}{}^{\kappa}
 + C_{\nu\sigma\alpha}{}^{\kappa}\xi_{\mu\kappa}{}^{\alpha}
 -  4 \xi_{\mu\kappa[\nu}L_{\sigma]}{}^{\kappa}  + 6g_{\mu[\nu} \xi^{\kappa}{}_{\sigma\alpha]}L_{\kappa}{}^{\alpha}
 \nonumber
\\
 && + 8L_{\alpha\kappa} \nabla_{[\nu}C_{\sigma]}{}^{\alpha}{}_{\mu}{}^{\kappa}  +  \frac{1}{6} R \xi_{\mu\nu\sigma}
  - \frac{1}{3}  C_{\nu\sigma\mu}{}^{\kappa}\nabla_{\kappa} R
  \;,
 \label{weylwave5}
\\
  R^{(H)}_{\mu\nu}[g] &=& 2L_{\mu\nu} + \frac{1}{6} R g_{\mu\nu}
 \label{weylwave6red}
 \;.
\end{eqnarray}
Here $R^{(H)}_{\mu\nu}[g] $ is defined as
\begin{equation}
 R^{(H)}_{\mu\nu} := R_{\mu\nu} - g_{\sigma(\mu}\hat\nabla_{\nu)} H^{ \sigma}
  \;.
 \label{ricci_riccired}
\end{equation}
Further, the field $\xi_{\mu\nu\sigma}$ above will, in the final space-time, be the Cotton tensor, related to the Schouten tensor $L_{\mu\nu}$ as
\begin{eqnarray*}
 \xi_{\mu\nu\sigma}
\,=\, 4\nabla_{[\sigma}L_{\nu]\mu}
 \,=\, 2 \nabla_{[\sigma}R_{\nu]\mu} +\frac{1}{3}g_{\mu[\sigma} \nabla_{\nu]} R
 \;.
\end{eqnarray*}
Finally, the operator  $ \Box_g^{(H)}$ is defined as
\begin{eqnarray}
 \nonumber
 \lefteqn{
 \Box_g^{(H)}v_{\alpha_1\dots\alpha_n}  :=
   \Box_g v_{\alpha_1\dots\alpha_n}
     - \sum_i g_{\sigma[\alpha_i}(\hat\nabla_{\mu]} H^{ \sigma})v_{\alpha_1\dots}{}^{\mu}{}_{\dots\alpha_n}
     \phantom{xxxxxxxxx}}
     &&
 \\
 && \phantom{xxxxxxxxxxx} +\sum_i(2L_{\mu\alpha_i} -R^{(H)}_{\mu\alpha_i} + \frac{1}{6}Rg_{\mu\alpha_i})v_{\alpha_1\dots}{}^{\mu}{}_{\dots\alpha_n}
 %\\
% \nonumber
%  && \phantom{xxxxxxxxxxx} + \sum_i g^{\sigma[\beta_i}(\hat\nabla_{\sigma} H^{ \mu]})v_{\alpha_1\dots\alpha_n}{}^{\beta_1\dots}{}_{\mu}{}^{\dots\beta_m}
% \\
%  &&\phantom{xxxxxxxxxxx} -\sum_i(2L^{\mu\beta_i} -R_{(H)}^{\mu\beta_i} + \frac{1}{6}Rg^{\mu\beta_i})v_{\alpha_1\dots\alpha_n}{}^{\beta_1\dots}{}_{\mu}{}^{\dots\beta_m}
%
\;,\phantom{xx }
 \label{5VI13.1}
\end{eqnarray}
with $   \Box_g  = \nabla^\mu \nabla_\mu$,   where in the sums in \eq{5VI13.1} the index $\mu$ occurs as the $i$'th index on $v_{\alpha_1\ldots \alpha_n}$.

Some comments concerning \eq{5VI13.1} are in order. First,  if $g$ solves Friedrich's equations \eq{mconf1}-\eq{mconf6} in the gauge $H^\sigma=0$, then  $ \Box_g^{(H)}= \Box_g$, so one may wonder why we are not simply using $\Box_g$. The issue is that the operator $ \Box_g$ on tensor fields of nonzero valence  contains second-order derivatives of the metric, so that the principal part of a system of equations obtained by replacing  $ \Box_g^{(H)}$  by $ \Box_g $ in \eq{weylwave1}-\eq{weylwave6red} will \emph{not} be diagonal. This could be cured by adding  equations obtained by differentiating \eq{weylwave6red}, which is not convenient as it leads to further constraints. Instead, one observes~\cite[Section~3.1]{TimConformal}
that the second derivatives of the metric appearing in  $ \Box_g $ can be eliminated in terms of the remaining fields above. For example, for a covector field $v$,
\begin{eqnarray*}
 \Box_g v_{\lambda} &=& g^{\mu\nu}\partial_{\mu}\partial_{\nu}v_{\lambda}-g^{\mu\nu}(\partial_{\mu}\Gamma^{\sigma}_{\nu\lambda})v_{\sigma} +f_{\lambda}(g,\partial g, v, \partial v)
 \\
 &=& g^{\mu\nu}\partial_{\mu}\partial_{\nu}v_{\lambda}
  +(R_{\lambda}{}^{\sigma} - \partial_{\lambda}(g^{\mu\nu}\Gamma^{\sigma}_{\mu\nu}))v_{\sigma} +f_{\lambda}(g,\partial g, v, \partial v)
  \\
 &=& g^{\mu\nu}\partial_{\mu}\partial_{\nu}v_{\lambda}
  +(R_{\lambda}{}^{\sigma} - \partial_{\lambda}H^{\sigma})v_{\sigma} +f_{\lambda}(g,\partial g, v, \partial v,\hat g, \partial\hat g, \partial^2 \hat g )
   \\
 &=& g^{\mu\nu}\partial_{\mu}\partial_{\nu}v_{\lambda}
  +(R_{\mu\lambda}^{(H)} + g_{\sigma[\lambda}\hat\nabla_{\mu]} H^{ \sigma})v^{\mu} +f_{\lambda}(g,\partial g, v, \partial v,\hat g, \partial\hat g, \partial^2 \hat g )
   \;,
\end{eqnarray*}
with $f_\lambda$ changing from line to line.
This leads to the definition
\begin{eqnarray}
 \Box_g^{(H)}v_{\lambda} &:=& \Box_g v_{\lambda} - g_{\sigma[\lambda}(\hat\nabla_{\mu]} H^{ \sigma})v^{\mu}
  +(2L_{\mu\lambda} -R^{(H)}_{\mu\lambda} + \frac{1}{6}Rg_{\mu\lambda})v^{\mu}
   \;,
   \phantom{xxxx}
\end{eqnarray}
consistently with \eq{5VI13.1}.

An identical calculation  shows that the operator \eq{5VI13.1} has the properties just
described  for higher-valence covariant tensor fields.

It follows from the above that
the principal part of $ \Box_g^{(H)}$ is $g^{\mu\nu}\partial_\mu\partial_\nu$. This implies that the principal part of \eq{weylwave1}-\eq{weylwave6red} is diagonal, with principal symbol equal to  $g^{\mu\nu}p_\mu p_\nu$ times the identity matrix. In particular, we can use~\cite{Dossa97} to find solutions of our equations whenever suitably regular initial data are at disposal.

Let $(x^0,x^1\equiv r,x^A)$ be coordinates adapted to the light-cone $C_{i^-}$ of $i^-$ as in~\cite[Section~4]{CCM2}, and let $\kappa$ measure how the coordinate $ x^1$ differs from an affine parameter along the generators of the light-cone of $i^-$:
$$
 \nabla_1 \partial_1|_{C_{i^-}} = \kappa \partial_1
 \;.
$$
%%
%We denote by $D$ the covariant derivative of the ($r$-dependent family of) metrics $\tilde g := \ol g_{AB}dx^A dx^B$ on $S^2$.

There are various gauge freedoms in the equations above. To get rid of this we can, and will, impose
\bel{2VI13.1}
  \hat g_{\mu\nu} = \eta_{\mu\nu}
  \;,
  \
  R=0
  \;,
  \
  H^\sigma = 0
  \;,
  \
  \kappa=0
  \;,
  \
   s|_{C_{i^-}} = -2
  \;.
\ee
The condition $ \hat g_{\mu\nu} = \eta_{\mu\nu}$ is a matter of choice. The conditions $R=0$ and $H^\sigma=0$ are classical, and can be realized by solving wave equations. The condition $\kappa=0$ is a choice of parameterization of the generators of $C_{i^-}$. The fact that $ s  $ can be made a negative constant on $C_{i^-}$ is justified in  Appendix~\ref{A2VI13.1}, see Remark~\ref{R5VI13.1}. As already pointed out, the value $s=-2$ is a matter of convenience, and can be achieved by a  constant rescaling of $\Theta$ and of the field $d_{\alpha\beta\gamma}{}^\delta$.

Consider the set of fields
\bel{2VI13.2}
 \Psi =(g_{\mu\nu},  L_{\mu\nu}, C_{\mu\nu\sigma}{}^{\rho}, \xi_{\mu\nu\sigma}, \Theta,  s)
 \;.
\ee
We will denote by
\bel{30V13.1}
\mathring \Psi:= (\mathring g_{\mu\nu}, \mathring L_{\mu\nu}, \mathring C_{\mu\nu\sigma}{}^{\rho}, \mathring\xi_{\mu\nu\sigma},\mathring \Theta , \mathring s  )
%\;,
\ee
the (characteristic) initial data for $\Psi$
defined along $C_{i^-}$.

Set
\bel{7X12.2}
 \omega_{AB}\equiv \breve{\mathring L}_{AB}:=   {\mathring L}_{AB} - \frac 12 \mathring g^{CD}{\mathring L}_{CD} \mathring g_{AB}
 \;,
\ee
and define $\lambda_{AB}$ to be the solution of the equation
   \bel{7X12.1}  (\partial_1- r^{-1})\lambda_{AB} = - 2 \omega_{AB}
   \ee
satisfying $ \lambdahere_{AB}=O(r^3)$.%
\footnote{When $\mathring L_{AB}$ arises from the restriction to the light-cone of a bounded space-time tensor, it holds that $\omega_{AB}=O(r^2)$ or better. We will only consider such initial data here, then there exists  a unique solution of \eq{7X12.1} satisfying  $\lambdahere_{AB}=O(r^3)$ .}
The following can be derived~\cite[Sections~4.2, 4.3 \& 6.4]{TimConformal}
from \eq{mconf1}-\eq{mconf6} and the gauge conditions \eq{2VI13.1}:
\begin{eqnarray}
\mathring g_{\mu\nu} &= &\eta_{\mu\nu} \;,
 \label{xi_constraints1}
\\
\mathring L_{1\mu} &=& 0\;, \quad
 \mathring L_{0A} \,=\, \frac{1}{2}\mathrm{D}^B\lambdahere_{AB}\;, \quad
  \mathring g^{AB} \mathring L_{AB} \,=\,0\;,
 \label{xi_constraints1b}
\\
 \mathring C_{\mu\nu\sigma\rho}&=&0\;,
 \label{xi_constraints2}
 \\
 \mathring\xi_{11A} &=& 0
  \,,
 \\
 \mathring\xi_{A1B} &=& -2r\partial_1( r^{-1}\omega_{AB})
 \label{xi_constraints3}
  \;,
 \\
 \mathring\xi_{ABC} &=& 4 \mathrm{D}_{[C}\omega_{B]A} - 4r^{-1}\mathring g_{A[B} \mathring L_{C]0}
   \;,
\\
 \mathring \xi_{01A} &=& \mathring g^{BC}\mathring \xi_{BAC}
  \;,
\\
\partial_1 \mathring\xi_{00A} &=&
%\frac{1}{4}\tilde\nabla^B(\lambda_A{}^C\omega_{BC}) - \frac{1}{2}\omega_{A}{}^B\mathring L_{0B} - \frac{1}{4}\lambda_B{}^C\tilde\nabla^B\omega_{AC}
\mathrm{D}^B(\lambda_{[A}{}^C\omega_{B]C})
-2 \mathrm{D}^B\mathrm{D}_{[A}\mathring L_{B]0}
  +\frac{1}{2}\mathrm{D}^B\overline \xi_{A1B}
 \nonumber
\\
&& - 2r\mathrm{D}_A \rho
 + r^{-1}\mathring \xi_{01A}  + \lambda_A{}^B\mathring \xi_{01B} \;,
 \label{xi_constraints6}
\\
 \mathring\xi_{A0B}  &=&  \lambda_{[A}{}^C\omega_{B]C} - 2\mathrm{D}_{[A}\mathring L_{B]0} + 2r\mathring g_{AB}\rho  - \frac{1}{2}\mathring\xi_{A1B} \;,
\\
   4(\partial_1+ r^{-1})\mathring L_{00}  &=& \lambda^{AB} \omega_{AB} - 2\mathrm{D}^A \mathring L_{0A}  -4r \rho
\;,
 \label{xi_constraints9}
\end{eqnarray}
with $\mathring\xi_{00A}=O(r)$, $\mathring L_{00}=O(1)$, and where $\rho$ is the unique bounded solution of
\begin{eqnarray}
 (\partial_1+\ 3r^{-1})\rho &=& \frac{1}{2}r^{-1}\mathrm{D}^A\partial_1\mathring L_{0A} - \frac{1}{4}\lambda^{AB}\partial_1(r^{-1}\omega_{AB})
\label{xi_constraints10}
 \;.
\end{eqnarray}
Here, and elsewhere, the symbol $\mathrm{D}_A$ denotes the covariant derivative of $\mathring g_{AB}\mathrm{d}x^A\mathrm{d}x^B$.

Let $s_{AB}$ denote  the unit round metric on $S^2$.
We will need the following result~\cite[Theorem~6.5]{TimConformal}:

\begin{theorem}
\label{T13X12.5}
Consider a set of smooth fields $\Psi$
defined in a neighborhood $\mcU$ of $i^-$ and satisfying \eq{weylwave1}-\eq{weylwave6red} in $I^+(i^-)$. Define  the data  \eq{30V13.1}
by restriction of $\Psi$ to $C_{i^-}$, suppose that $\mathring \Theta=0$ {and $\mathring s=-2$}.
Then the fields
$$
 ( g_{\mu\nu},  L_{\mu\nu},  d_{\mu\nu\sigma}{}^{\rho} = \Theta^{-1}C_{\mu\nu\sigma}{}^{\rho}, \Theta, s )
$$
solve on $\mcD^+(\dot J^+(i^-)\cap \mcU)$ the conformal field equations \eq{mconf1}-\eq{mconf6} in the gauge
 \eq{2VI13.1}, with the conformal factor $\Theta$   positive on $I^+(i^-)$  sufficiently close to $i^-$,  with  $\mathrm{d}\Theta\ne 0 $ on $C_{i^-}\setminus \{i^-\}$
near  $i^-$, and with $\mathring C_{\mu\nu\sigma}{}^{\rho}=0$,
if and only if \eq{7X12.1}-\eq{xi_constraints10} hold with $\rho$ and $r^{-3}\lambda_{AB}$ bounded.
\qed
\end{theorem}

\begin{Remark}
\label{r6.12.1}
It follows from  \eq{xi_constraints3} that a necessary condition for existence of solutions as in the theorem is $\omega_{AB}=O(r^3)$.
\end{Remark}

\begin{Remark}
\label{r6.12.3}
Note that solutions of the ODEs \eq{xi_constraints6} and \eq{xi_constraints9}  are rendered unique by the conditions
$\mathring\xi_{00A}=O(r)$ and  $L_{00}=O(1)$, which follow from regularity of the fields at the vertex.
\end{Remark}

We use overlining to denote \emph{restriction to the $\eta$-light-cone} of $i^-$.

Consider a set of fields $(\tilde L_{\mu\nu}, \tilde\xi_{\mu\nu\rho})$ defined in a neighborhood of $ i^-$, and set
\bel{13X12.1}
  \omega_{AB}:=\olt{ L}_{AB} - \frac 12 \olt g{}^{CD}\olt{ L}_{CD} \olt g_{AB}
  \;.
% \;,
\ee
%     and $\omega_{AB}$ defined on $C_{i^-}$.
We will say that $(\tilde L_{\mu\nu}, \tilde\xi_{\mu\nu\rho})$  provides an \emph{approximate  solution of the constraint equations } if \eq{7X12.1}-\eq{xi_constraints10}  hold up to $\orinf$ error terms. Thus it must hold that
\begin{eqnarray}
\label{14X12.10}
&&\hspace{-8em} \olt L_{1\mu}=\orinf\;, \quad
 \olt L_{0A} = \frac{1}{2}\nottmathrmD^B\lambdahere_{AB} + \orinf\;, \quad
  \olt g^{AB} \olt L_{AB} =\orinf\;,
   \phantom{xx}
%\\
% \olt C_{\mu\nu\sigma}{}^{ \rho}&=&\orinf\;,
 \\
 \olt\xi_{11A} &=& \orinf
\label{14X12.11}
  \,,
 \\
 \olt\xi_{A1B} &=& -2r\partial_1( r^{-1}\omega_{AB})+ \orinf
\label{14X12.12}
  \;,
 \\
 \olt\xi_{ABC} &=& 4\nottmathrmD_{[C}\omega_{B]A} - 4r^{-1}\olt g_{A[B} \olt L_{C]0} + \orinf
\label{14X12.13}
   \;,
\\
 \olt \xi_{01A} &=& \olt g^{BC}\olt \xi_{BAC}+ \orinf
\label{14X12.14}
  \;,
\\
\partial_1 \olt\xi_{00A} &=&\nottmathrmD^B(\lambda_{[A}{}^C\omega_{B]C})
- 2\nottmathrmD^B\nottmathrmD_{[A}\olt L_{B]0}  +\frac{1}{2}\nottmathrmD^B\overline \xi_{A1B}
 \nonumber
\\
&&- 2r\nottmathrmD_A \tilde \rho
 + r^{-1}\olt \xi_{01A}  + \lambda_A{}^B\olt \xi_{01B}+ \orinf
\label{14X12.15}
  \;,
\\
 \olt\xi_{A0B}  &=&  \lambda_{[A}{}^C\omega_{B]C} - 2\nottmathrmD_{[A}\olt L_{B]0}
   + 2r\olt g_{AB}\tilde \rho
  \nonumber
\\
 && - \frac{1}{2}\olt\xi_{A1B}+ \orinf
\label{14X12.16}
  \;,
\\
   4(\partial_1+ r^{-1})\olt L_{00}  &=& \lambda^{AB} \omega_{AB}  -2\nottmathrmD^A \olt L_{0A}  -4r \tilde \rho + \orinf
\;,
\label{14X12.17}
\end{eqnarray}
with $\olt\xi_{00A}=O(r)$, $\olt L_{00}=O(1)$,
where $\tilde \rho$ is a bounded solution of
\begin{eqnarray}
 (\partial_1+\ 3r^{-1})\tilde \rho = \frac{1}{2}r^{-1}\nottmathrmD^A\partial_1\olt L_{0A} - \frac{1}{4}\lambda^{AB}\partial_1(r^{-1}\omega_{AB}) + \orinf
 \;,
  \label{14X12.18}
\end{eqnarray}
and where $\lambda_{AB}$ is the solution of \eq{7X12.1}  satisfying
$ \lambdahere_{AB}=O(r^3)$, or differs from that solution by $\orinf$ terms.

Our first main result is the following:

\begin{theorem}
\label{t13X12.1}
Let $\tilde g_{\mu\nu}$ be a smooth metric defined near $i^-$ such that for small $r$ we have
$$
%\ol{\tilde R}=\orinf\;,\
\ol{\tilde  g_{\mu\nu}-\eta_{\mu\nu}}=\orinf
%\;,\
%\ol{ \tilde H}{}^\sigma=\orinf\;,\ \ol{\tilde\kappa}=\orinf\;,\ \ol{\tilde s}+2=\orinf
 \;.
$$
Let $\tilde L_{\mu\nu}$ be the Schouten tensor of $\tilde g_{\mu\nu}$, let $\tilde \xi_{\alpha\beta\gamma}$ be the Cotton tensor of $\tilde g_{\mu\nu}$ and let $\tilde C_{\alpha\beta\gamma\beta}$ be its Weyl tensor.
Assume that $(\tilde L_{\mu\nu}, \tilde\xi_{\mu\nu\rho})$  solves the approximate constraint equations.

Then  there exist  smooth fields $
 ( g_{\mu\nu},  L_{\mu\nu},   C_{\mu\nu\sigma}{}^{\rho}, \Theta, s )
$ defined in a   neighbourhood of $i^-$ such that the fields
$$
 ( g_{\mu\nu},  L_{\mu\nu},  d_{\mu\nu\sigma}{}^{\rho} = \Theta^{-1}C_{\mu\nu\sigma}{}^{\rho}, \Theta, s )
$$
 solve the conformal field equations \eq{mconf1}-\eq{mconf6} in $I^+(i^-)$, satisfy the gauge conditions \eq{2VI13.1},
with
\bel{13X12.3}
    \ol \Theta = 0
 \;, \quad
  \ol C_{\mu\nu\sigma}{}^{\rho} =0
 \;, \quad
 \breve{\ol  L}_{AB} =\omega_{AB}
 \;,
\ee
with the conformal factor $\Theta$   positive on $I^+(i^-)$  sufficiently close to $i^-$, and  with  $\mathrm{d}\Theta\ne 0 $ on $C_{i^-}\setminus \{i^-\}$
near  $i^-$.
%with  $\ol g_{\mu\nu}=\ol \eta_{\mu\nu}$.
\end{theorem}

\proof
We will apply Theorem~\ref{T13X12.5} to a suitable evolution of the initial data. For this we need to correct  $\Psi$ by   smooth fields so that the restriction to the light-cone of the new $ \Psi$
 satisfies the constraint equations as needed for that theorem. Subsequently, we define new fields
\beaa
 &
  \check g_{\mu\nu}= \tilde g_{\mu\nu} + \delta g_{\mu\nu}\;,
  \quad
  \check L_{\mu\nu}= \tilde  L_{\mu\nu} + \delta  L_{\mu\nu}\;,%
%   &
%\\
%    &
\quad
    \check\xi_{\mu\nu\sigma}= \tilde \xi_{\mu\nu\sigma} + \delta \xi_{\mu\nu\sigma}
    \;,
    &
\eeaa
as follows:

We let $\delta   g_{\mu\nu}$ be any smooth tensor field defined in a neighborhood of $i^-$ which is $   \oyinf$ and which
satisfies
$$
  \overline{ \delta g}{}_{\mu\nu} =  \overline{\eta_{\mu\nu} - \tilde g_{\mu\nu}}
  \;.
$$
Indeed, it follows from e.g. \cite[Equations (C4)-(C5)]{ChConeExistence} that the $y$--coordinates components $ { \old g_{\mu\nu}}$ of $g$ are $\orinf$,
and  existence of their smooth extensions follows from~\cite[Lemma~A.1]{ChJezierskiCIVP}.
This extension procedure will be used extensively from now on without further reference.

 %\ptcr{rewrites and rearrangements until the end of the proof}
To continue,
  we let $\delta \xi_{\alpha\beta\gamma}$ be any smooth tensor defined in a neighborhood of $i^-$, with $y$-coordinate-components    which are $\oyinf $, such that
  \begin{enumerate}
    \item
$\old \xi_{11A} =-\olt \xi_{11A} $;
 \item $
  \old \xi_{A1B} =-\olt \xi_{A1B}  -2r\partial_1( r^{-1}\omega_{AB})
 % \;,
  $
(recall that $\omega_{AB}$ has been defined in \eq{13X12.1});

\item
  $
  \old \xi_{ABC} =- \olt\xi_{ABC} + 4\nottmathrmD_{[C}\omega_{B]A} - 4r^{-1}\olt g_{A[B} \olt L_{C]0}
  \;;
  $
\item
  $
  \old\xi_{01A} =- \olt \xi_{01A} + \olt g^{BC}\olt \xi_{BAC}
  \;;
  $
\item
  $\old \xi_{00A}$ is the solution vanishing at $r=0$ of the system of ODEs
\bean
\partial_1 (\old \xi_{00A} + \olt\xi_{00A} )  &=&\nottmathrmD^B(\lambda_{[A}{}^C\omega_{B]C})
- 2\nottmathrmD^B\nottmathrmD_{[A}\olc L_{B]0}  +\frac{1}{2}\nottmathrmD^B\olc \xi_{A1B}
 \nonumber
\\
&&- 2 r\nottmathrmD_A \check \rho
 + r^{-1}\olc \xi_{01A}  + \lambda_A{}^B\olc \xi_{01B}
  \;,
\eeal{13X12.4}
where $\check \rho$ is the unique bounded solution of
\begin{eqnarray}
 (\partial_1+\ 3r^{-1})\check \rho &=& \frac{1}{2}r^{-1}\nottmathrmD^A\partial_1\olc L_{0A} - \frac{1}{4}r^{-1}\lambda^{AB}\partial_1(r^{-1}\omega_{AB})
% \;.
\end{eqnarray}
(it follows from \cite[Appendix~B]{ChConeExistence} that $\old \xi_{00A}$ is $\orinf$);

\item
  $
  \old\xi_{A0B} = -\olt\xi_{A0B}  + \lambda_{[A}{}^C\omega_{B]C} - 2\nottmathrmD_{[A}\olt L_{B]0} + 2r\olt g_{AB}\check \rho  - \frac{1}{2}\olt\xi_{A1B}
  \;.
  $
\end{enumerate}

Finally, we let $  \delta L_{\mu\nu}$ be any smooth tensor field defined in a neighborhood of $i^-$, the $y$-components of which are $O(|y|^\infty)$, such that:
\begin{enumerate}
  \item  $ \overline {\delta L}_{1\mu}=-\olt L_{1\mu}$;
  \item $
  \overline{\delta L}_{0A} = -
 \olt L_{0A}+ \frac{1}{2}\nottmathrmD^B\lambdahere_{AB}
 \;;
 $
\item
 $\delta L_{AB}= f  \eta_{AB}$,
  where $f$ is any smooth function defined in a neighborhood of $i^-$   which is $\oyinf $ such that
 $$
  2 \ol f \equiv \ol\eta^{AB}\old L_{AB} = -
  \olt g^{AB} \olt L_{AB}
 $$
(we emphasise that the $\eta_{AB}$-trace-free part of $  L_{AB}$ coincides thus with the $\eta_{AB}$-trace-free part of $  \tilde L_{AB}$);
\item
 $\old L_{00}$ is the solution vanishing at $r=0$ of the system of ODEs
\bea
   4(\partial_1+ r^{-1})(\old L_{00}+\olt L_{00})  &=&  \lambda^{AB} \omega_{AB} - 2\nottmathrmD^A \olt L_{0A}  -4r \check \rho
%\;.
\end{eqnarray}
(it follows from \cite[Appendix~B]{ChConeExistence} that $\old L_{00}$ is $\orinf$).
\end{enumerate}

Let
$$(g_{\mu\nu},  L_{\mu\nu}, C_{\mu\nu\sigma}{}^{\rho}, \xi_{\mu\nu\sigma}, \Theta,  s)$$
be a solution of  \eq{weylwave1}-\eq{weylwave6red} with initial data
$$(\mathring g_{\mu\nu}, \mathring L_{\mu\nu}, \mathring C_{\mu\nu\sigma}{}^{\rho}, \mathring\xi_{\mu\nu\sigma},\mathring \Theta, \mathring s ):= \ol{ ( \eta_{\mu\nu},  \check L_{\mu\nu},0, \check\xi_{\mu\nu\sigma}, 0,  -2 )}
\;.
$$
%}
%
A solution exists by \cite[Th\'eor\`eme~2]{Dossa97}.
It follows by construction that the   hypotheses of Theorem~\ref{T13X12.5} hold, and the theorem is proved.
\qed\medskip

An alternative way of obtaining solutions of our problem proceeds via the following system of  conformal wave equations:
\begin{eqnarray}
 \Box^{(H)}_{ g} L_{\mu\nu}&=&  4 L_{\mu\kappa} L_{\nu}{}^{\kappa} -  g_{\mu\nu}| L|^2
  - 2\Theta d_{\mu\sigma\nu}{}^{\rho}  L_{\rho}{}^{\sigma}
 + \frac{1}{6}\nabla_{\mu}\nabla_{\nu} R
 \label{wave1}
  \;,
  \\
  \Box _{ g}  s  &=& \Theta| L|^2 -\frac{1}{6}\nabla_{\kappa} R\,\nabla^{\kappa}\Theta  - \frac{1}{6} s  R
 \label{wave2}
  \;,
  \\
  \Box _{ g}  \Theta &=& 4 s-\frac{1}{6} \Theta  R
 \label{wave3}
  \;,
  \\
  \Box^{(H)}_g d_{\mu\nu\sigma\rho}
  &=& \Theta d_{\mu\nu\kappa}{}^{\alpha}d_{\sigma\rho\alpha}{}^{\kappa}
   - 4\Theta d_{\sigma\kappa[\mu} {}^{\alpha}d_{\nu]\alpha\rho}{}^{\kappa}
  + \frac{1}{2}R d_{\mu\nu\sigma\rho}
 \label{wave4}
  \;,
  \\
  R^{(H)}_{\mu\nu}[g] &=& 2L_{\mu\nu} + \frac{1}{6} R g_{\mu\nu}
 \label{wave6}
  \;.
\end{eqnarray}

Any solution of the conformal field equations \eq{mconf1}-\eq{mconf6}
\bel{30V13.11}
(R=0,\hat g_{\mu\nu}=\eta_{\mu\nu},H^\mu = 0,\kappa=0,\mathring s=-2)
\ee
necessarily satisfies~\cite[Sections~4.2 \& 4.3]{TimConformal}: 
\begin{eqnarray}
 \label{14X12.1}
&&\hspace{-8em} \mathring g_{\mu\nu} = \eta_{\mu\nu} \;, \quad\mathring L_{1\mu}=0\;, \quad
 \mathring L_{0A} = \frac{1}{2}\nottmathrmD^B\lambdahere_{AB}\;, \quad
  \mathring g^{AB} \mathring L_{AB} =0\;,
\\
   \mathring d_{1A1B} &=&- \frac{1}{2}\partial_1(r^{-1}\omega _{AB})
 \label{14X12.1a}
 \;,
\\
    \mathring d_{011A} &=&  \frac{1}{2}r^{-1}\partial_1\mathring L_{0A}\;,
\\
 \mathring d_{01AB}&=&  r^{-1}\nottmathrmD_{[A}\mathring L_{B]0} -\frac 12 r^{-1}\lambda_{[A}{}^C \omega_{B]C}
 \;,
\\
 (\partial_1+\ 3r^{-1})\mathring d_{0101} &=& \nottmathrmD^A\mathring d_{011A} + \frac{1}{2}\lambdahere^{AB}\mathring d_{1A1B}   \;,
   \label{14X12.3}
\\
 2(\partial_1+ r^{-1})\mathring d_{010A}   &=&  \nottmathrmD^B( \mathring d_{01AB} - \mathring d_{1A1B} )+  \nottmathrmD_A \mathring d_{0101}
 +2 r^{-1} \mathring d_{011A}
  \nonumber
\\
&&
 + 2 \lambda_{A}{}^B \mathring d_{011B}\;,
  \label{14X12.4}
 \\
  4( \partial_{1} - r^{-1} )\breve {\mathring d}_{0A0B}
&=& (\partial_{1} - r^{-1}) {\mathring d}_{1A1B}  + 2(\nottmathrmD_{(A} \mathring d_{B)110})\,\breve{}
 +   4 (\nottmathrmD_{(A} \mathring d_{B)010}    )\breve{}
 \nonumber
\\
 &&
  + 3 \lambda_{(A}{}^C\mathring d_{B)C01}
   +3\mathring d_{0101}\lambdahere_{AB}
\;,
 \phantom{xxxx}
  \label{14X12.5}
\\
  4(\partial_1+ r^{-1})\mathring L_{00}  &=&   \lambdahere^{AB} \omega_{AB}- 2\nottmathrmD^A \mathring L_{0A}   -4r \mathring d_{0101}
\;.
 \label{14X12.6}
\end{eqnarray}

We have
the following result~\cite[Theorem~5.1]{TimConformal}:

\begin{theorem}
\label{main_result}
A smooth solution
$$(g_{\mu\nu},  L_{\mu\nu}, d_{\mu\nu\sigma}{}^{\rho}, \Theta,  s)$$
of the system  \eq{wave1}-\eq{wave6},
with initial data
$$
 (\mathring g_{\mu\nu}, \mathring L_{\mu\nu}, \mathring d_{\mu\nu\sigma}{}^{\rho}, \mathring \Theta=0, \mathring s=-2 )
% \;,
$$
on $C_{i^-}$, solves on $\mcD^+(\dot J^+(i^-))$  the conformal field equations \eq{mconf1}-\eq{mconf6}
in the gauge \eq{30V13.11},
with $\Theta$   positive on   $I^+(  {i^-})$  sufficiently close to $i^-$, and with $\mathrm{d}\Theta\ne 0 $ on $C_{i^-}\setminus \{i^-\}$
near  $i^-$, if
and only if \eq{14X12.1}-\eq{14X12.6} hold with
  $\omega_{AB}(r,x^A)$ and $\lambda_{AB}(r,x^A)$ defined by \eq{7X12.2}-\eq{7X12.1}.
\end{theorem}

\begin{Remark}
 \label{r6.12.2}
It follows from  \eq{14X12.1a} that a necessary condition for existence of solutions as in Theorem~\ref{main_result} is $\omega_{AB}=O(r^4)$.
Note that this is stronger than what is needed in Theorem~ \ref{t13X12.1}, see Remark~\ref{r6.12.1}. It would be of interest to clarify the question of existence of data needed for Theorem~\ref{T13X12.5} with $\omega_{AB}=O(r^3)$ properly.
\end{Remark}

\begin{Remark}
\label{r6.12.4}
Note that the solutions of the ODEs \eq{14X12.3}-\eq{14X12.6}  are rendered unique by the conditions
$\mathring d_{0101}=O(1)$, $\mathring d_{010A} =O(r)$, $\breve {\mathring d}_{0A0B}=O(r^2)$ and
$\mathring L_{00}=O(1)$, which follow from regularity of the fields at the vertex.
\end{Remark}

A smooth metric $\tilde g$
will be called an \emph{approximate solution of the constraint equations  \eq{14X12.1}-\eq{14X12.6}} if $
 \tilde C^{\alpha}{}_{\beta\gamma\delta}=\tilde \Theta \tilde d^{\alpha}{}_{\beta\gamma\delta}
$ for some smooth function $\tilde \Theta$ vanishing on $C_{i^-}$ and for some smooth tensor $  \tilde d^{\alpha}{}_{\beta\gamma\delta}
$, where $
 \tilde C^{\alpha}{}_{\beta\gamma\delta}
$  is the Weyl tensor of $\tilde g$, and if \eq{14X12.1}-\eq{14X12.6} hold on the light-cone of $i^-$ up to terms which are $\orinf$, where $\tilde L_{\mu\nu}$ is the Schouten tensor of $\tilde g$, and where $\omega_{AB}$ and $\lambda_{AB}$ are, possibly up to $\orinf$ terms, given by \eq{7X12.2}-\eq{7X12.1}.

Our second main result is the following:

\begin{theorem}
\label{t5VI13.1x}
Let $\tilde g_{\mu\nu}$ be
an approximate solution of the constraint equations   \eq{14X12.1}-\eq{14X12.6} defined near $i^-$.
Then  there exist  smooth fields
$$
 ( g_{\mu\nu},  L_{\mu\nu},  d_{\mu\nu\sigma}{}^{\rho} = \Theta^{-1}C_{\mu\nu\sigma}{}^{\rho}, \Theta, s )
$$
defined in a   neighbourhood of $i^-$ which solve the conformal field equations \eq{mconf1}-\eq{mconf6} in $I^+(i^-)$, satisfy the gauge conditions \eq{2VI13.1},
with
\bel{13X12.3b}
    \ol \Theta = 0
 \;, \quad
  \ol C_{\mu\nu\sigma}{}^{\rho} =0
 \;, \quad
 \breve{\ol  L}_{AB} =\omega_{AB}
 \;,
\ee
with the conformal factor $\Theta$   positive on $I^+(i^-)$  sufficiently close to $i^-$, and  with  $\mathrm{d}\Theta\ne 0 $ on $C_{i^-}\setminus \{i^-\}$
near  $i^-$.
%with  $\ol g_{\mu\nu}=\ol \eta_{\mu\nu}$.
\end{theorem}

\proof
We will apply Theorem~\ref{main_result} to a suitable evolution of the initial data. For this we need to correct  $( \tilde g_{\mu\nu}, \tilde  L_{\mu\nu} , \tilde  d_{\mu\nu\sigma\rho})$ by   smooth fields so that the new initial data on the light-cone
 satisfy the constraint equations as needed for that theorem. The construction of the new fields
\bea
 &
  \check g_{\mu\nu}= \tilde g_{\mu\nu} + \delta g_{\mu\nu}\;,
  \quad
  \check L_{\mu\nu}= \tilde  L_{\mu\nu} + \delta  L_{\mu\nu}\;,
\quad
    \check d_{\mu\nu\sigma\rho} = \tilde  d_{\mu\nu\sigma\rho} + \delta  d_{\mu\nu\sigma\rho}
    \;,
    \phantom{xxx}
    &
\eeal{5VI13.11}
is essentially identical to that of the new fields of the proof of Theorem~\ref{t13X12.1}, the reader should have no difficulties filling-in the details.
We emphasise that the trace-free part of  $\delta L_{AB}$ is chosen to be zero, hence the trace-free part of $\check L_{AB}$ coincides with the trace-free part of $\tilde L_{AB}$ on the light-cone.

Once the  fields \eq{5VI13.11} have been constructed, we let
$$(g_{\mu\nu},  L_{\mu\nu}, d_{\mu\nu\sigma\rho},   \Theta,  s)$$
be a solution of  \eq{wave1}-\eq{wave6} with initial data
$$(\mathring g_{\mu\nu}, \mathring L_{\mu\nu}, \mathring d_{\mu\nu\sigma\rho},  \mathring \Theta, \mathring s ):= \ol{ ( \eta_{\mu\nu},  \check L_{\mu\nu} , \check d_{\mu\nu\sigma\rho}, 0,  -2 )}
\;.
$$
%}
%
A solution exists by \cite[Th\'eor\`eme~2]{Dossa97}.
It follows by construction that the   hypotheses of Theorem~\ref{main_result} hold, and the theorem is proved.
\qed\medskip
%
%\ptcr{a lot of material went to recycling.tex}

\section{Proof of Theorem~\ref{T19XII12.1}}
 \label{s30V13.1}

We are ready now to prove Theorem~\ref{T19XII12.1}: Let  $\varsigma$ be the cone-smooth  tensor field of the statement of the theorem. Thus, there exists a smooth tensor field $\tilde d_{\alpha\beta\gamma\delta}$  with the algebraic symmetries of the Weyl tensor so that
$\varsigma $ is the pull-back of
\bel{6VI13.31}
 \tilde d_{\alpha\beta\gamma\delta}\ell^\alpha \ell^\gamma
\ee
to $C_O\setminus \{O\}$.

Let $\tilde \psi_{MNPQ}$
be a totally-symmetric two-index spinor associated to $\tilde d_{\alpha\beta\gamma\delta}$ in the usual way~\cite[Section~3]{FriedrichNullData} (compare~\cite{PenroseRindler84v1}). Set $\theta^0=dt $, $\theta^1 = dr$.
Let $\gamma$ be a generator of $C_{0}$, and let  $\theta^2$, $\theta^3$ be a pair of
covector fields  so that $\{\theta^\mu\}$ forms an orthonormal basis of $ T^* \mcM $ over $\gamma$ and which are $\eta$-parallel propagated along $\gamma$.
Then $\varsigma$ can be written as
$$\varsigma= \varsigma_{ab}\theta^a \theta^ b
\;,
$$
%.
with $a,b$ running over $\{2,3\}$.
 %\ptcr{added}
By construction, the coordinate components $\varsigma_{AB}$ of $\varsigma$ coincide with  the coordinate components $\ol{\tilde d}_{1A1B}$ of the restriction to the light-cone of $ {\tilde d}_{1A1B}$, and thus define a unique field $\omega_{AB}$ by integrating \eq{14X12.1a} with the boundary condition  $\omega_{AB}=O(r^2)$.

Let the basis $\{e _\mu\}$ be dual to $\{ \theta^\mu\}$, set $m=e_2+\sqrt{-1} e_3$. Then the radiation field $\psi_0$ of~\cite[Equation~(5.3)]{FriedrichNullData},  defined using $\tilde \psi_{MNPQ}$, equals
$$
 \psi_0=\varsigma_{ab}m^a m^b
 \;.
$$
%.
(Under a rotation of $\{e_3,e_4\}$ the field $\psi_0$ changes by a phase, and defines thus a section of a spin-weighted bundle over $C_O\setminus\{O\}$.)
Conversely, any radiation field $\psi_0$ of \cite{FriedrichNullData} arises from a unique cone-smooth $\varsigma_{ab}$ as above.

It has been shown in~\cite[Propositions~8.1 \& 9.1]{FriedrichNullData} that the radiation field $\psi_0$, hence $\varsigma$,
defines a smooth Lorentzian metric $\ttilde g$ such that the resulting collection of fields $(\ttilde g_{\mu\nu},  \ttilde L_{\mu\nu}, \ttilde d_{\mu\nu\sigma \rho},  \ttilde \Theta,  \ttilde s)$ satisfies
\eq{mconf1}-\eq{mconf6} up to error terms which are $\oyinf$, with $\ttilde g_{\mu\nu}|_{C_{O}}= \eta_{\mu\nu}$.
The construction in~\cite{FriedrichNullData} is such that the field $ \varsigma$ calculated from the  field  $\tilde d_{\alpha\beta\gamma\delta}$ of \eq{6VI13.31}
coincides with the field $ \varsigma$ calculated from the  field  $\ttilde d_{\alpha\beta\gamma\delta}$ associated with the metric $\ttilde g$. Hence the  fields $\omega_{AB}$ associated with $\tilde d_{\alpha\beta\gamma\delta}$ and $\ttilde d_{\alpha\beta\gamma\delta}$ are identical.
The conclusion follows now from Theorem~\ref{t5VI13.1x}.
% \ptcr{hypotheses verified?}
%
\qed

\section{Alternative data at $\scrim$}
 \label{s31V13.1}

Recall that there are many alternative ways to specify initial data for the Cauchy problem for the vacuum
Einstein equations on a (usual) light-cone, cf.\ e.g.~\cite{ChPaetz}. Similarly there are many ways to provide initial data on a light-cone emanating from past timelike infinity. In Theorem~\ref{T19XII12.1} some components of the rescaled Weyl tensor $d_{\mu\nu\sigma\rho}$ have been prescribed as free data. As made clear in the proof of that theorem,
this is equivalent to providing some components of the rescaled Weyl spinor $\psi_{MNPQ}$, providing thus an alternative equivalent prescription. Our Theorems~\ref{t13X12.1} and \ref{t5VI13.1x}
use instead the components \eq{13X12.1} of the rescaled Schouten tensor $\tilde L_{\mu\nu}$. These components are related directly to the free data of Theorem~\ref{T19XII12.1} via the constraint equation \eq{14X12.1a}.
It is clear that further possibilities exist. Which of these descriptions of the degrees of freedom of the gravitational field at large retarded times is  most useful for physical applications remains to be seen.

\appendix
\section{The $\ol s=-2$ gauge}
 \label{A2VI13.1}

We start with some terminology. We say that a
function $f$ defined on a space-time neighbourhood of the
origin is $o_m(|y|^k)$ if $f$ is $C^m$ and if for $0\le \ell
\le m$ we have
$$
 \lim_{|y|\to 0}
 |y|^{\ell-k}\partial_{\mu_1} \ldots
 \partial_{\mu_\ell} f =0
 \;,
$$
%,
where $|y|:=\sqrt{\sum_{\mu=0}^n (y^\mu)^2}$.

A similar definition will be used for functions defined in a
neighbourhood of $O$ on the future light-cone
$$
 C_O=\{y^0 = |\vec y|\}
 \;.
$$
For this, we parameterize
$C_O$ by  coordinates $\vec y = (y^i)\in {\R }^n$, and  we say that
a function $f$ defined on a  neighbourhood of $O$ within $C_O$
is $o_m(r^k)$ if $f$ is a $C^m$ function   of the coordinates
$y^i$ and if for $0\le \ell \le m$ we have $\lim_{r\to 0}
r^{\ell-k}\partial_{\mu_1} \ldots
\partial_{\mu_\ell} f =0$, where
$$r:= |\vec y|\equiv \sqrt{\sum_{i=1}^n (y^i)^2}
 \;.
$$
%.
We further set
$$
   \Theta^i:= \frac{y^i} r
 \;.
$$

A function $\varphi$ defined on $C_O$ will be said to be \emph{$C^k$-cone-smooth} if there exists a function  $f$  on space-time of differentiability class $C^k$ such that $\varphi$ is the restriction of $f$ to $C_O$. We will simply say \emph{cone-smooth} if $k=\infty$.

The following lemma will be used repeatedly:

\begin{lemma}[Lemma~A.1 in \cite{ChJezierskiCIVP}]
 \label{L23I.1}
 Let $k\in \N$.
A function $\varphi$ defined on a light-cone $C_{O}$ is the
trace $\overline{f}$ on $C_{O}$ of a $C^{k}$ space-time function
$f$ if and only if $\varphi$ admits an expansion of the form
\begin{equation}
 \label{23XI.1}
\varphi= \sum_{p=0}^{k}f_{p}r^{p}+o_k(r^{k})
 \;,
\end{equation}
with
\begin{equation}
 \label{6XI.1}
f_{p}\equiv
f_{i_{1}\ldots i_{p}}\Theta ^{i_{1}}\cdots \Theta ^{i_{p}}
+
f^{\prime}{}_{i_{1}\ldots i_{p-1}}\Theta ^{i_{1}}\cdots \Theta ^{i_{p-1}}
 \;,
\end{equation}
where $f_{i_{1}\ldots i_{p}}$ and $f_{i_{1}\ldots
i_{p-1}}^{\prime }$ are numbers.

The claim remains true with $k=\infty$ if
 \eqref{23XI.1} holds for all $k\in \N$.
  \qed
\end{lemma}

Coefficients $f_p$ of the form \eq{6XI.1} will be said to be \emph{admissible}.

  \medskip

One of the elements needed for the construction in~\cite{FriedrichNullData} is provided by the following result:

\begin{Proposition}
 \label{P29VIII12.1}
Let $p$ be a point in a smooth space-time $(\mcM, g )$, $\mcU $ a   neighborhood of $p$, and   $S_{\mu \nu}[g]$ the trace free part of the
Ricci tensor of $g$. Let  $\ell^{\nu} $ denote  the field of null directions tangent to $\partial J^+(p) \cap \mcU  $. Let $a > 0$ be a real number and let  $\beta$ be a one-form at $p$. Then, replacing $\mcU $ by a smaller neighborhood of $p$ if necessary,  there exists a unique smooth function $  \theta$ defined on $  \mcU  $   satisfying
\bel{29VIII12.21}
 \mbox{$ {\theta} =a$ and $d \theta = \beta$  at $p$, \quad  $\ell^{\mu} \ell ^{\nu} S_{\mu \nu}[\theta^2 g] = 0$
on $\partial J^+(p) \cap \mcU  $,}
\ee
and such that
\bel{29VIII12.22}
 \mbox{the Ricci scalar of $\theta^2 g$ vanishes on $  J^+(p) \cap \mcU  $.}
\ee
\end{Proposition}

\begin{Remark}
  \label{R5VI13.1}
  {\rm
  It follows from \eq{mconf4} multiplied by $\nabla^\mu \Theta$ that $s$ is constant on  $\partial J^+(p) \cap \mcU  $  when the gauge \eq{29VIII12.21} has been chosen.
  }
\end{Remark}

\proof
In dimension $n$ let  $g'=\phi^{4/(n-2)}g$, then
$$
 R'_{\mu\nu}=R_{\mu\nu}- 2\phi^{-1}\nabla_{\mu}\nabla_{\nu} \phi+\frac{2n}{n-2}
 \phi^{-2}\nabla_{\mu}\phi\nabla_{\nu} \phi-\frac{2}{n-2}
 \phi^{-1}(\nabla^{\sigma}\nabla_{\sigma}\phi+\phi^{-1}|d\phi|^2)g_{\mu\nu}
 \;.
$$
%
%$$
%R'\phi^{(n+2)/(n-2)}=-\frac{4(n-1)}{n-2}\nabla^k\nabla_k\phi+R\phi\;.
%$$
%%
So, in dimension $n=4$, and with $\phi=\theta$ we obtain
%\tim{greek indices}
%
$$
S'_{\mu\nu}=S_{\mu\nu}- 2\theta^{-1}\left (\nabla_{\mu}\nabla_{\nu} \theta- \frac 1 4 \Delta \theta g_{\mu\nu}\right)+ 4
\theta^{-2} \left( \nabla_{\mu}\theta\nabla_{\nu} \theta - \frac 14 |\nabla \theta|^2 g_{\mu\nu}\right)
 \;.
$$
We overline restrictions of space-time functions to the light-cone.
The equation $\overline{\ell^{\mu} \ell ^{\nu} S_{\mu \nu}[\theta^2 g]} = 0$ takes thus the form
% \ptcr{2 changed to $\frac 12$ at several places; please crosscheck}
%
$$
\overline{\ell^{\mu} \ell ^{\nu} \nabla_\mu\nabla_\nu \theta  - 2
\theta^{-1}( \ell^{\mu}   \nabla_\mu\theta)^2}=
 \frac 12\overline {\theta \ell^\mu \ell^\nu S_{\mu\nu} }
 \;.
$$
In coordinates adapted to the light-cone as in \cite[Appendix~A]{CCM2}, so that $\ell^\mu \partial_\mu = \partial_r$, with $r$ an affine parameter, this reads
$$
\ol\theta^{-1}\partial_r^2 \ol\theta  - 2
\ol\theta^{-2}(\partial_r\ol\theta)^2=
 \frac 12 \ol S_{11}
 \;.
$$
Setting
$$
 \varphi:= \frac{\partial_r\ol \theta}{\ol \theta}
 \;,
$$
this can be rewritten as
\bel{29VIII12.11}
 \partial_r \varphi  = \varphi^2 +
 \frac 12 \ol S_{11}
 \;.
\ee

It is useful to introduce some notation.
% \ptcr{moved here}
As in \cite{CCM2}, we underline the components of a tensor in the coordinates $y^\mu$, thus:
$$
 \underline{S_{\mu\nu}}=S\Big(\frac{\partial}{\partial y^\mu},\frac{\partial}{\partial y^\mu}\Big)
  \;,
  \qquad
  {S_{\mu\nu}}=S\Big(\frac{\partial}{\partial x^\mu},\frac{\partial}{\partial x^\mu}\Big)
  \;,
$$
etc, where $(x^\mu):=(y^0 - |\vec y|, |\vec y|, x^A)$, with $x^A$ being any local coordinates on $S^2$. We write interchangeably $x^1$ and $r$.

The initial data for $\varphi$ are
\bel{29VIII12.11b}
  \varphi (0) =\frac 1 a ( \underline{ \beta_0} + \underline{ \beta_i} \Theta^i)
 \;.
\ee

To show that $\ol\theta$ is cone-smooth, it suffices to prove that
$$
 \psi:=r\varphi
$$
is $C^k$-cone-smooth for all $k$, as follows immediately from the expansions of Lemma~\ref{L23I.1}, together with integration term-by-term in the formula
$$
 \ln \left(\frac{\ol \theta} a\right)  =  \int_0^r \varphi
 \;,
$$
compare \cite[Lemma~B.1]{ChConeExistence}.

We shall proceed by induction. So suppose that  $\psi$ is $C^k$-cone-smooth. The result is true for $k=0$ since
every solution of \eq{29VIII12.11} is continuous in all variables, and $r\varphi$  tends to zero as $r$ tends to zero, uniformly in $\Theta^i\in S^{2}$.

It follows that the source term $\overline S_{11}$ in \eq{29VIII12.11} can be written as
$$
\ol S_{11}=r^{-2}\ol{\underbrace{\underline{r^2 S_{00}}+2t\underline{S_{0i}}y^{i}
           +\underline{S_{ij}}y^{i}y^{j}}_{=:\chi}}:= r^{-2} \ol\chi
           \;,
$$
where $\chi$ is a smooth function on space-time. We thus have
\bel{29VIII12.11c}
 \partial_r \varphi  = \varphi^2 +
 \frac 12\ol S_{11} = r^{-2}(\psi^2 + \frac{1}{2} \ol \chi)
 \;.
\ee
The function $\psi $ is $C^k$-cone-smooth and $O(r )$, and can thus be written in the form \eq{23XI.1}-\eq{6XI.1},
$$
 \psi  = \sum_{p=1}^{k}f_{p}r^{p}+o_k(r^{k})
 \;.
$$
Squaring we obtain
$$
 \psi^2  = \sum_{p=2}^{k+1}f'_{p}r^{p}+o_k(r^{k+1})
 \;,
$$
for some new admissible coefficients $f'_p$.
The function $ \ol \chi$ is $C^{k+1}$-cone-smooth and $O(r^2)$, and can thus be written in the form \eq{23XI.1}-\eq{6XI.1} with $k$ replaced by $k+1$ there,
$$
 \ol\chi = \sum_{p=2}^{k+1}f''_{p}r^{p}+o_{k+1}(r^{k+1})
 \;.
$$
Hence
\bel{29VIII12.13xx}
 \partial_r \varphi  =  \sum_{p=2}^{k-1}f'''_{p}r^{p-2}+o_k(r^{k-1 })
 \;
\ee
for some admissible coefficients $f'''_p$.
Integration gives
\bel{29VIII12.13}
  \varphi  =  \frac 1 a ( \underline{ \beta_0} + \underline{ \beta_i} \Theta^i) + \sum_{p=2}^{k }\frac 1 {p-1} f'''_{p}r^{p-1}+o_k(r^{k })
 \;,
\ee
and thus
\bel{29VIII12.13x}
  \psi = r \varphi  =  \frac 1 a ( \underline{ \beta_0}r + \underline{ \beta_i} y^i) + \sum_{p=2}^{k+1}\frac 1 {p-1} f'''_{p}r^{p }+o_k(r^{k+1})
 \;.
\ee
Differentiating $r\varphi$ with respect to $r$  and using \eq{29VIII12.13xx}
we further obtain
\bel{29VIII12.13a}
  \partial_r\psi = \partial_r\left( \frac 1 a ( \underline{ \beta_0}r + \underline{ \beta_i} y^i) + \sum_{p=2}^{k+1}\frac 1 {p-1} f'''_{p}r^{p }\right)+o_k(r^{k})
 \;.
\ee
Let $X= X^A\partial_A$ be any vector field on $S^2$, then $X(\varphi)$ solves the equation obtained by differentiating \eq{29VIII12.11c},
\bel{29VIII12.14}
 \partial_r X(\varphi)   = 2 \varphi X(\varphi) +
\frac 12 X( \ol S_{11})
 \;.
\ee
Equivalently,
\beaa
 X(r\varphi)(r,x^A)  & = &
   e^{2 \int_0^r   \varphi(\tilde r,x^A)  d\tilde r }\Big(X(r\varphi(0, x^A))
\\
   && + \frac 12 r\int_0^r e^{-2 \int_0^{\hat r}   \varphi(\tilde r,x^A)  d\tilde r } X( \ol S_{11})(\hat r, x^A) d\hat r
 \Big)\;.
\eeaa
The right-hand side is $C^k$-cone-smooth. We conclude that $\psi$ is $C^{k+1}$-cone-smooth. This finishes the induction, and proves that $\ol \theta$ is cone-smooth.

The existence and uniqueness of a solution $\theta$ of \eq{29VIII12.22} which equals $\ol \theta$ on $C_O$ follows now from \cite[Th\'eor\`eme~2]{Dossa97}.
\qed
\medskip
%

%\ptcr{comment out what follows to process the remainder of the notes}

%\bibliographystyle{amsplain}
%\bibliography{../references/hip_bib,%
%../references/reffile,%
%../references/newbiblio,%
%../references/newbiblio2,%
%../references/chrusciel,%
%../references/bibl,%
%../references/howard,%
%../references/bartnik,%
%../references/myGR,%
%../references/newbib,%
%../references/Energy,%
%../references/dp-BAMS,%
%../references/prop2,%
%../references/besse2,%
%../references/netbiblio,%
%../references/PDE}
%
%\end{document}

%\ptcr{the previous recycling material went to recycling.tex, a beginning of some arguments in schema.tex}
%\input{schema}
%\input{recycling}

%\noindent{\sc Acknowledgements} PTC acknowledges useful discussions with

\noindent {\sc Acknowledgements} Useful discussions with Helmut Friedrich are acknowledged.

\bibliographystyle{amsplain}
\bibliography{../references/hip_bib,%
../references/reffile,%
../references/newbiblio,%
../references/newbiblio2,%
../references/chrusciel,%
../references/bibl,%
../references/howard,%
../references/bartnik,%
../references/myGR,%
../references/newbib,%
../references/Energy,%
../references/dp-BAMS,%
../references/prop2,%
../references/besse2,%
../references/netbiblio,%
../references/PDE}

\end{document}